# Comprendre les archives : vers de nouvelles interfaces de recherche reposant sur l'annotation sémantique des documents

*Understanding Archives : Towards New Research Interfaces Relying on the Semantic Annotation of Documents*

**Nicolas GUTEHRLÉ (1), Iana ATANASSOVA (1, 2)**

(1)
Université de Franche-Comté, CRIT
F-25000 Besançon, France
nicolas.gutehrle@univ-fcomte.fr

(2) Institut Universitaire de France (IUF), France
iana.atanassova@univ-fcomte.fr

**Résumé**. Les campagnes de numérisation menées ces dernières années par les bibliothèques et archives ont facilité l'accès aux documents de leurs collections. Cependant, explorer et exploiter ces contenus restent des tâches difficiles en raison de la quantité de documents consultables. Dans cet article, nous montrons comment l'annotation sémantique du contenu textuel de corpus d'études issus de documents d'archives permet d'en faciliter l'exploitation et la valorisation. Nous présentons d'abord un cadre méthodologique pour la construction de nouvelles interfaces qui reposent sur la sémantique textuelle, puis abordons les verrous technologiques actuels, et leurs potentielles solutions. Nous terminons par la présentation d'un cas pratique de l'application de ce cadre.
**Mots-clés**. Archives, valorisation, gestion des connaissances, ouverture et accessibilité des archives, découvrabilité des documents, annotation sémantique, interfaces de recherche

**Abstract.** The digitisation campaigns carried out by libraries and archives in recent years have facilitated access to documents in their collections. However, exploring and exploiting these documents remain difficult tasks due to the sheer quantity of documents available for consultation. In this article, we show how the semantic annotation of the textual content of study corpora of archival documents allow to facilitate their exploitation and valorisation. First, we present a methodological framework for the construction of new interfaces based on textual semantics, then address the current technological obstacles and their potential solutions. We conclude by presenting a practical case of the application of this framework

1. **Introduction**

Ces dernières années, les bibliothèques et archives ont mené de nombreuses campagnes de numérisation de leurs collections. Si ces campagnes ont facilité l'ouverture et l'accessibilité des documents d'archives à un public plus large, leur découvrabilité et la valorisation de leurs contenus restent des tâches difficiles. En effet, les documents numérisés sont habituellement peu structurés. En y accédant via des moteurs de recherche, l'utilisateur est limité par les requêtes reposant sur des mots-clés, et est alors submergé par l'abondance de documents consultables mais pas toujours pertinents pour sa recherche. Ainsi, il est nécessaire de structurer le contenu textuel des documents d'archives afin d'améliorer les interfaces de recherches existantes et de faciliter l'exploration, l'exploitation et la valorisation des « données massives du passé » (Kaplan et Di Lenardo, 2017).

Cette problématique relève du domaine des méthodes de Traitement Automatique des Langues (TAL), telles que la reconnaissance d'entités nommées, la modélisation de sujet (*Topic Modelling*), la détection d'événements ou l'analyse de sentiments. Ces méthodes viennent enrichir les documents par l'ajout d'annotations sémantiques. L'application de ces méthodes aux documents d'archives a suscité un intérêt important ces dernières années, comme en témoignent les nombreux travaux publiés (Boros et al., 2022; Dominguès *et al.*, 2019; Ehrmann, 2008; Ehrmann *et al.*, 2021; Ide et Woolner, 2004; Marjanen *et al.*, 2020; Smith, 2002; Sprugnoli *et al.*, 2016; Yang *et al.*, 2011), ainsi que l'apparition de journaux dédiés, tels que *Journal of Data Mining and Digital Humanities* (JDMDH) ou *Journal on Computing and Cultural Heritage* (JOCCH), d'ateliers de conférence comme *Language Technology for Cultural Heritage, Social Sciences, and Humanities* (LaTeCH) ou *HistoInformatics*, de hackathons tels que *Helsinki Digital Humanities Hackathon*, de compétitions comme *Identifying Historical People, Places and other Entities* (HIPE) et de réseaux internationaux comme *Common Language Resources and Technology Infrastructure* (CLARIN).

La structuration et l'annotation des documents contribuent à la valorisation des archives de deux manières. Dans un premier temps, les annotations sémantiques obtenues par les méthodes de TAL sont exploitables par les interfaces de recherche pour indexer les documents, comme dans les plateformes *impresso*[1], *NewsEye*[2] ou *Retronews*[3]. Cette indexation fine permet des requêtes basées sur les annotations et sont plus précises par rapport à celles qui n'utilisent que des mots clés. Cette approche limite le nombre de documents retournés et facilite leur découvrabilité. Par exemple, un utilisateur pourrait spécifier les mots-clés et thématiques qu'il ou elle souhaite retrouver dans les documents retournés, ou bien filtrer les résultats par rapport à une entité nommée en particulier. Dans un second temps, les annotations sémantiques peuvent être exploitées pour générer des visualisations telles que des diagrammes, des nuages de mots, des cartes ou des frises chronologiques afin de permettre une lecture distante des données (Moretti, 2013). Ces méthodes peuvent faciliter l'étude de questions de recherche en histoire comme l'ont montré des projets internationaux tels que *Trading Consequences* (Klein *et al.*, 2014), *Living With Machines*[4] ou *Viral Texts*[5]. En cela, l'application des méthodes de TAL aux documents d'archives s'inscrit dans le contexte des Humanités Numériques.

---

[1] https://impresso-project.ch/

[2] https://www.newseye.eu/

[3] https://www.retronews.fr/

[4] https://livingwithmachines.ac.uk/

[5] https://viraltexts.org/

**Comprendre les archives : vers de nouvelles interfaces de recherche reposant sur l'annotation sémantique des documents**

La structuration des documents d'archives par les méthodes de TAL implique cependant des verrous technologiques et méthodologiques qu'il est nécessaire de considérer. Tout d'abord, la qualité des résultats obtenus par ces méthodes est impactée par des facteurs tels que la qualité de la transcription des textes ou la structure des documents. Cet impact est d'autant plus important que les outils et ressources disponibles ont été construits initialement pour le traitement de documents modernes. L'adaptation de ces outils aux documents d'archives nécessite des ressources qui sont rares. La fiabilité de ces outils peut également être questionnée, en particulier pour les outils reposant sur des algorithmes de *machine* et *deep learning*. En effet, si les résultats obtenus par ces méthodes sont prometteurs, leurs fonctionnement est opaque et difficilement interprétable. Enfin, la structuration des fonds documentaires ne doit pas se faire au détriment des méthodes de recherche des utilisateurs.

Dans cet article, nous proposons un cadre méthodologique modulaire pour la construction de nouvelles interfaces de recherche qui repose sur l'annotation sémantique d'un corpus d'étude de documents d'archives dans le but d'en faciliter l'exploitation et la valorisation. Nous décrivons les différentes étapes de ce cadre et faisons également l'état de l'art des outils disponibles pour les traitements. Nous détaillons en particulier les possibles verrous technologiques et méthodologiques, et leurs potentielles solutions. Enfin, nous présentons un cas pratique de l'application de ce cadre méthodologique. Le reste de cet article se présente comme suit : nous présentons tout d'abord le cadre méthodologique en section 2. Les verrous ainsi que le cas d'étude sont décrits dans les section 3 et 4. Enfin, nous présentons notre conclusion et des pistes pour des travaux futurs dans la section 5.

## 2. Cadre méthodologique

Dans cette section, nous présentons un cadre méthodologique, qui consiste en trois grandes étapes : une première étape de pré-traitement des documents, une seconde d'annotation de leur structure sémantique, et une dernière d'augmentation des interfaces de recherche actuelles. Ces étapes sont représentées par la figure 1.

**Figure 1** : *Etapes du cadre méthodologique*

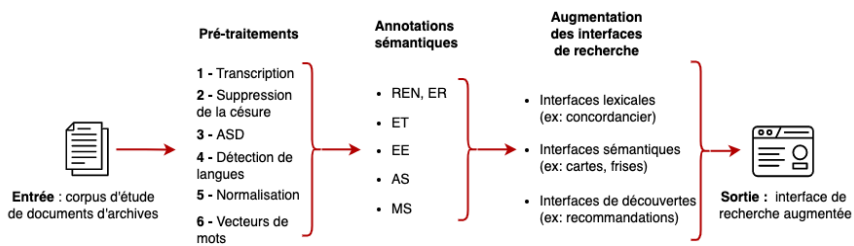

### 2.1. Pré-traitements

L'étape de pré-traitement comprend une série de traitements visant à rendre les documents exploitables par les méthodes de TAL. Tout d'abord, il est nécessaire d'en extraire le contenu textuel. Pour cela, deux types de solutions existent : on peut transcrire les documents de manière participative (*crowdsourcing*) à l'aide de plateformes de transcription, telle que Scripto[6], comme cela a été fait pour les projets *Old Weather*[7] ou *What's on the Menu*[8]. Une autre alternative est d'employer des outils de reconnaissance optique de caractères (*Optical Character Recognition*, OCR), tels que ABBYY FineReader[9] ou Tesseract[10], ou de reconnaissance d'écriture manuscrite (*Handwritten Text Recognition,* HTR) tels que Transkribus[11] ou eScriptorium[12].

Une seconde étape importante de pré-traitement du texte est la suppression des césures. En typographie, la césure consiste à diviser un mot par un trait d'union afin qu'il respecte la justification du texte. La césure peut ainsi introduire des artefacts, notamment si la transcription du texte est obtenue via des méthodes d'OCR. Pour la supprimer et corriger les mots séparés, on peut employer des règles de traitement simples (Généreux et Spano, 2015).

Une troisième étape consiste à procéder à une analyse de la structure des documents (ASD) (*Document Layout Analysis*). Elle se divise en deux étapes : l'analyse de la structure physique (ASP) (*Physical Layout Analysis*), qui identifie les régions textuelles du document, et l'analyse de la structure logique (ASL) (*Logical Layout Analysis*), qui en identifie les éléments logiques (titre, en-tête, notes de bas de pages, tableau, paragraphes, phrases). Le rôle de ces analyses intervient à plusieurs niveaux : l'identification des régions textuelles permet de situer les mots sur les pages des documents numérisés. Ainsi les termes employés lors d'une recherche par mots-clés peuvent-ils être surlignés directement sur l'image du document. L'ASL permet, quant à elle, de rechercher un type de contenu textuel en particulier, et notamment d'appliquer des traitements successifs sélectionnés selon les différents types de contenu. Par exemple, l'identification des en-têtes des pages est utile pour en extraire des métadonnées, tandis que l'identification des paragraphes peut servir, entre autres, à trouver les entités nommées qui y sont mentionnés. Certains logiciels d'OCR comme ABBYY FineReader peuvent procéder à l'ASP. Ils produisent ainsi des fichiers dans des formats spécifiques tels que le XML ALTO, qui contiennent à la fois le contenu textuel et la structure physique du document. Ces logiciels ne peuvent cependant pas procéder à l'ASL. Pour cela, il faut employer des systèmes développés sur mesure qui exploitent, par exemple, les documents XML ALTO (Atanassova et Gutehrlé, 2022) ou l'image des documents (Barman *et al.,* 2021).

Les fonds documentaires peuvent également être plurilingues. Dans de tels cas, une quatrième étape de pré-traitement de détection de la langue (Majumder *et al.*, 2002) des contenus textuels est nécessaire afin de pouvoir employer les outils de TAL adaptés à chaque langue. De plus, suivant l'époque d'origine des documents d'archives, la ou les langues dans lesquelles ces documents sont écrits peuvent être plus ou moins éloignées de leur forme moderne. Cette variation peut ainsi avoir un impact sur les résultats obtenus tant par les moteurs de recherche que par les méthodes de TAL. Ainsi, une cinquième étape de pré-traitement est de normaliser les documents, de telle sorte que la langue utilisée soit la plus proche possible de la langue moderne. On peut traiter ces variations par l'emploi de méthodes à base de règles (Piotrowski, 2012) ou d'algorithmes de *machine* et *deep learning* (Bollmann et

---

[6] https://scripto.org/

[7] https://www.oldweather.org/

[8] https://menus.nypl.org/

[9] https://pdf.abbyy.com/

[10] https://github.com/tesseract-ocr/tesseract

[11] https://readcoop.eu/transkribus/

[12] https://www.escriptorium.uk/index.html

**Comprendre les archives : vers de nouvelles interfaces de recherche reposant sur l'annotation sémantique des documents**

Søgaard, 2016; Domingo et Nolla, 2018) qui font correspondre les termes anciens à leurs équivalents modernes.

Enfin, une dernière étape possible consiste à obtenir des vecteurs de mots (*word embeddings*) depuis le corpus d'étude. Les vecteurs de mots reposent sur l'hypothèse distributionnelle (Firth, 1957; Harris, 1954), qui stipule que le sens d'un mot est défini par son contexte. Les informations syntaxiques et sémantiques des mots sont appris à partir d'un corpus par des algorithmes tels que *word2vec* (Mikolov *et al.*, 2013) ou *fastText* (Bojanowski *et al.*, 2016), et encodés sous forme de vecteurs denses (Almeida et Xexéo, 2019). Les vecteurs de mots peuvent servir dans de nombreuses tâches comme le calcul de la similarité entre documents, la suggestion de mots-clés ou l'entraînement de modèles de *machine* ou *deep-learning*.

### 2.2. Annotation sémantique

L'étape d'annotation consiste à structurer le corpus d'étude par l'ajout d'annotations sémantiques. Nous désignons par annotation sémantique l'annotation de contenus textuels selon des catégories sémantiques prédéfinies. On peut notamment évoquer les méthodes d'annotation suivantes : la reconnaissance d'entités nommées (REN) qui consiste à détecter toute expression faisant référence à une entité réelle, telle qu'une personne, un lieu ou une organisation (Ehrmann, 2008; Ehrmann *et al.*, 2021) ; l'extraction de relation (ER) pour identifier les relations entre entités nommées (Klein *et al.*, 2014; Plum *et al.*, 2022) ; la détection d'expressions temporelles (ET) pour identifier l'expression de la temporalité explicite (« 1917 ») comme implicite (« hier ») ; l'extraction d'événements (EE), de leur date d'occurrence et des personnes qui y ont participées (Boros *et al.*, 2022; Ide et Woolner, 2004; Smith, 2002) ; l'analyse de sentiments (AS) pour mesurer la polarité (positif, négatif, neutre) ou les émotions exprimées dans les textes (Dominguès *et al.*, 2019; Sprugnoli *et al.*, 2016) ; la modélisation de sujet (MJ) pour identifier les thématiques des documents (Marjanen *et al.*, 2020; Yang *et al.*, 2011).

De nombreux outils sont aujourd'hui disponibles pour appliquer de telles méthodes. Les outils GATE (Cunningham, 2002), NLTK (Bird et Loper, 2004), spaCy[13] ou Stanza (Qi *et al.*, 2020) permettent de créer des modèles de classification pour des tâches comme la reconnaissance d'entités nommées, l'extraction de relation, l'analyse de sentiments ou la détection d'événements. Ces outils sont hybrides : ils permettent de créer des systèmes à base de règles comme des systèmes reposant sur des algorithmes de *machine* et *deep-learning*. Les outils MALLET (McCallum, 2002) et OpenNLP[14] quant à eux permettent d'employer des algorithmes de *machine-learning* pour des tâches de classification, tandis que les outils gensim (Řehůřek et Sojka, 2010) et MALLET permettent d'employer de modèles de modélisation de sujets et d'obtenir des vecteurs de mots. Les outils HeidelTime (Strötgen et Gertz, 2010) et SUTime (Chang et Manning, 2012) détectent les expressions temporelles et les normalisent, c'est-à-dire les convertissent dans un format informatique. Enfin l'outil HuggingFace[15] met à disposition les modèles récents reposant sur les techniques de *deep-learning* tels que BERT (Devlin *et al.*, 2019) ou GPT (Brown *et al.*, 2020) employés dans de nombreuses tâches.

### 2.3. Augmentation des interfaces de recherche

Au delà de leur utilisation dans l'indexation des documents, les annotations sémantiques peuvent également être exploitées pour concevoir de nouvelles interfaces de recherche qui compléteraient les interfaces de recherche actuelles. Ces interfaces permettraient d'aborder le corpus sous différents angles (Jatowt, 2021) et d'en avoir une lecture distante tout en conservant le lien avec les documents d'origine pour pouvoir revenir à une lecture proche. Nous envisageons trois types distincts d'interfaces : lexicales, sémantiques et de découvertes.

---

[13] https://spacy.io/

[14] https://opennlp.apache.org/

[15] https://huggingface.co/

Les interfaces lexicales permettraient d'étudier le vocabulaire du corpus. On pourrait ainsi observer les occurrences de termes dans le temps via un graphique, à la manière de Google N-gram Viewer[16], en contexte à l'aide d'un concordancier (Gutehrlé *et al.*, 2021) ou visualiser leur polarité sous la forme d'un nuage de mots (Sprugnoli *et al.*, 2016). De même, une interface pourrait exploiter les vecteurs de mots obtenus à partir du corpus afin d'en étudier les champs sémantiques.

Les interfaces sémantiques quant à elles exploiteraient les annotations sémantiques des documents. Une interface pourrait présenter pour chaque entité nommée identifiée une fiche contenant les informations qui lui sont propres, comme sa fréquence d'apparition ou les documents qui la mentionne, à la manière de l'application *Social Networks and Archival Context* (SNAC)[17]. Des méthodes de génération de texte pourraient rédiger automatiquement la biographie d'une entité à partir des informations extraites. Les relations entre entités pourraient être visualisées sous la forme d'un réseau à partir des résultats de l'extraction de relations. Les lieux identifiés pourraient être visualisés sur des cartes générées automatiquement (Blevins, 2014; Dominguès *et al.*, 2019; Gutehrlé *et al.*, 2021; Moncla et al., 2019). Pour cela, ces entités doivent être associées à des coordonnées géographiques à l'aide de gazetiers (*gazetteer*) (McDonough *et al.*, 2019) tels que Geonames[18] ou Pleiades (Bagnall *et al.*, 2016) lors d'une étape de géocodage (*geocoding*). De même, des frises chronologiques pourraient être générées automatiquement à partir des événements et expressions temporelles détectés (Gutehrlé *et al.*, 2022; Pasquali *et al.*, 2019). De plus, ces interfaces pourraient être augmentées par des informations contextuelles issues de bases de connaissances comme Wikidata. Par exemple, une entité nommée peut être associée à une page Wikipedia correspondante via une étape de résolution des entités nommées (*Named Entity Linking*) (Linhares Pontes *et al.*, 2019), tandis que les cartes et frises chronologiques peuvent être complétées par des événements importants tels que des batailles ayant eu lieux pendant la période traitée (Gutehrlé *et al.*, 2021). Un exemple d'interface sémantique est présenté par la figure 2.

Enfin, les interfaces de découvertes exploiteraient le contenu textuel des documents, les requêtes de l'utilisateur ainsi que les annotations sémantiques, pour rendre possible la sérendipité, c'est-à-dire le fait de faire une découverte fructueuse par hasard (Hyvönen 2020; Hyvönen et Rantala, 2019). Ce type d'interface reposerait tout d'abord sur l'emploi de systèmes de recommandation, qui suggéreraient à un utilisateur des contenus pouvant l'intéresser en se basant sur ses recherches actuelles et précédentes (Aktas *et al.*, 2004; Ricci *et al.*, 2010). De même, ces interfaces pourraient exploiter les annotations sémantiques, en particulier les relations entre entités nommées, pour suggérer des contenus associés aux documents consultés par l'utilisateur. Par exemple, le système pourrait suggérer des documents mentionnant les membres de la famille d'une entité mentionnée dans le document actuellement consulté. Enfin, une autre interface pourrait exploiter les vecteurs de mots pour assister l'utilisateur dans l'écriture de sa requête en suggérant des termes sémantiquement proches ou plus adaptés au corpus (Ehrmann, Romanello, Clematide, *et al.*, 2020).

---

[16] https://books.google.com/ngrams/

[17] https://snaccooperative.org/

[18] https://www.geonames.org/

**Figure 2** : *Exemple d'une interface sémantique : carte générée automatiquement à partir des lieux identifiés dans « Le Matin » (1913-1915) (Gutehrlé et al., 2021)*

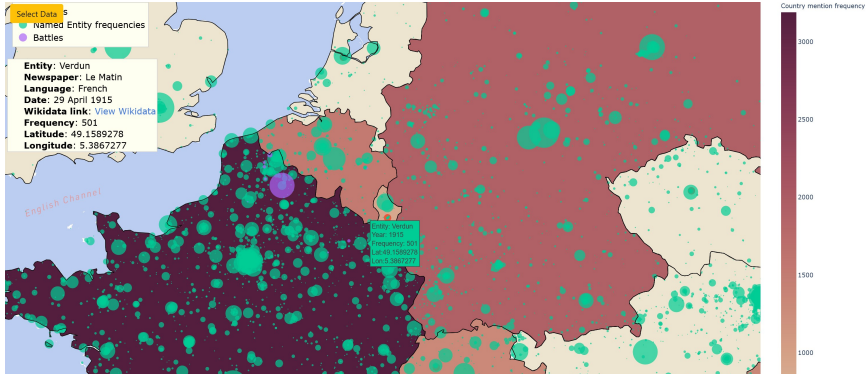

## 3. Verrous technologiques et méthodologiques

L'application de ce cadre rencontre cependant plusieurs difficultés et verrous technologiques comme méthodologiques, tant au niveau des documents eux-mêmes qu'au niveau de l'emploi des ressources et outils disponibles actuellement.

Une première difficulté est l'obtention des transcriptions des contenus textuels des documents. La mise en place de campagnes participatives de transcription nécessite tout d'abord de trouver des volontaires, de concevoir des guides d'annotation clairs et de faire appel à des experts pour contrôler la qualité des transcriptions obtenues. Ainsi, la mise en place de telles campagnes peut ne pas être possible ou adaptée à tous les projets. La transcription automatique des documents via des outils OCR implique également plusieurs difficultés. Tout d'abord, la qualité de ces transcriptions est très variable et dépend de nombreux facteurs tels que la qualité du document d'origine et de son double numérique, la mise en page, l'utilisation de typographies complexes, l'écriture manuscrite ou l'écart de la langue des documents avec la langue pour laquelle l'OCR a été préparé. Ces erreurs de transcriptions ne sont pas anodines, puisqu'elles ont un impact sur la qualité des résultats obtenus par les outils de TAL et les moteurs de recherche (Boros *et al.*, 2022; Ehrmann *et al.*, 2021; Linhares Pontes *et al.*, 2019). Une étape de correction des transcriptions OCR est donc nécessaire. Une première solution est la correction manuelle des transcriptions via des campagnes participatives, comme réalisé par le *Australian Newspapers Digitisation Program* (Holley, 2008). Une autre solution est d'employer des méthodes semi-automatiques qui détectent les erreurs et suggèrent des corrections à l'utilisateur, ou entièrement automatiques qui détectent et corrigent les erreurs à l'aide de règles ou de méthodes statistiques. Plus récemment, les méthodes de traduction neuronale ont été employées pour traduire le texte d'une mauvaise vers une bonne transcription (Molla et Cassidy, 2017; Rigaud *et al.*, 2019).

Une seconde problématique est soulevée par l'emploi de ressources externes telles que Wikidata ou Geonames pour augmenter les annotations et interfaces. En effet, la plupart de ces ressources ont été créées à partir de données contemporaines et peuvent donc ne pas correspondre à l'époque du corpus d'étude. Par exemple, les frontières d'un pays ou bien le nom d'un lieu peuvent être différents. Ces ressources restent de plus incomplètes. Ainsi, leur emploi induit le risque d'introduire des anachronismes et biais dans l'interprétation des résultats (Gutehrlé *et al.*, 2021).

Cette problématique est également présente dans l'utilisation des outils de TAL. En effet, la majorité des outils disponibles aujourd'hui ont été conçus pour traiter des documents écrits dans l'état moderne de la langue. De plus, ces fonds documentaires sont habituellement hétérogènes et contiennent des documents de natures et origines différentes (périodiques religieux, littérature, journaux orientés politiquement…) entre lesquels les langues et les styles d'écriture divergent. Les performances de ces outils sont donc moindres lorsqu'ils sont appliqués à des documents écrits dans plusieurs langues, dans un autre style ou dans un état plus ancien de la langue (Ehrmann *et al.*, 2016). Ceci soulève la question de l'adaptabilité des outils de TAL aux documents historiques (Sporleder, 2010). Une première solution est de normaliser la langue des documents de telle sorte à ce qu'elle corresponde à une état adapté à ces outils, comme suggéré en section 2.1. Cependant, plus l'état de la langue est éloigné de la langue moderne et plus la normalisation des variations orthographiques est difficile. Elle l'est d'autant plus pour les documents manuscrits anciens, où l'écriture est moins standardisée. De plus, ces méthodes se concentrent exclusivement sur la correspondance entre les graphies anciennes et modernes d'un mot, mais ne traitent pas leur évolution sémantique. Il y a donc un risque de faire correspondre des termes dont le sens a changé depuis. Ainsi, bien que cette étape facilite les requêtes et l'emploi d'outils de TAL, il est important de s'assurer qu'elle n'introduit pas d'artefacts.

Une autre possibilité est d'adapter les outils reposant sur des techniques de *deep learning* à de nouveaux corpus et nouveaux styles d'écriture à l'aide de méthode de *transfer learning* (Boros *et al.*, 2020). Le *transfer learning,* comme l'entraînement de modèles, requiert des jeux de données importants adaptés à la tâche traitée. S'il existe aujourd'hui de telles ressources dédiées aux documents d'archives (Ehrmann, Romanello, Flückiger, *et al.*, 2020; Gutehrlé et Atanassova, 2021; Hamdi *et al.*, 2021), ces dernières restent cependant rares. La création de telles ressources est une tâche longue et coûteuse (Nguyen *et al.*, 2021). Une manière d'accélérer la création de ces jeux de données est de mettre en place des campagnes participatives d'annotation. Une autre méthode consiste à employer des systèmes plus simples tels que des systèmes à base de règles pour pré-annoter les documents.

L'interprétabilité et la fiabilité des outils reposant sur les algorithmes de *machine* et *deep-learning* est également questionnée. En effet, leur fonctionnement interne et les processus qui mènent à des décisions sont complexes et difficilement interprétables, même par leurs créateurs. Ces systèmes sont ainsi souvent qualifiés de boîte-noires (Barredo Arrieta *et al.*, 2020). Cette opacité est problématique si l'étude et l'interprétation du corpus d'étude doit reposer sur les résultats obtenus par ces méthodes (Davis *et al.*, 2020). Une possible solution est tout d'abord d'employer des systèmes à base de règles ou des modèles plus simples tels des algorithmes de régression linéaire ou des arbres de décisions qui sont plus faciles à interpréter. Cependant, si l'on doit faire appel à des algorithmes plus complexes tels que des réseaux de neurones ou des modèles de langues comme BERT ou GPT, on peut faire appel à des algorithmes tels que LIME (Ribeiro *et al.*, 2016) qui assistent dans la compréhension des processus de ces systèmes.

Plusieurs travaux (Bahde, 2017; Beer et al., 2009) montrent l'intérêt des utilisateurs pour de telles nouvelles interfaces, en particulier pour assister dans la compréhension de la structure des fonds et identifier les liens entre les documents. Cependant, ces travaux montrent qu'un certain temps est nécessaire pour que les utilisateurs s'adaptent à ces outils et les intègrent à leurs méthodes de recherche. L'intérêt de ces outils diminue à mesure que l'utilisateur se familiarise avec les documents du fond étudié. Cet intérêt diminue d'autant plus pour les utilisateurs experts dans la recherche d'information dans les fonds documentaires. Ainsi, l'application de ce cadre méthodologique et la création de nouvelles interfaces de recherche permet la mise en place de nouvelles approches des fonds documentaire et l'accélération de la découverte de documents, mais ne vient en aucun cas remplacer les outils et méthodes de recherches actuels. De plus, il est impératif que les annotations générées automatiquement et les processus qui y mènent soient explicites et compréhensibles par les utilisateurs. Cependant, les sorties de certaines méthodes actuelles, telles que la modélisation de sujets ou la génération automatique de frises chronologiques, restent à être interprétées par l'utilisateur, ce qui requiert une expertise et limite l'utilité de ces outils auprès d'un public large. Ainsi, évaluer les méthodes de TAL selon des métriques telles que la Précision ou le Rappel ne suffit pas. Il est donc nécessaire d'évaluer ces outils et interfaces selon leur utilité pour les utilisateurs, par exemple à l'aide de questionnaires de satisfaction ou en identifiant les outils les plus fréquemment employés.

## 4. Etude de cas : corpus du projet EMONTAL

Dans cette section, nous présentons une application partielle du cadre méthodologique pour exploiter et valoriser un corpus de périodiques hétérogènes écrits en français et publiés au XXème siècle dans les régions Bourgogne et Franche-Comté. Cette application fait l'objet d'un travail en cours réalisé dans le cadre du projet Extraction et Modélisation Ontologique des Acteurs et Lieux pour la valorisation du patrimoine de Bourgogne Franche-Comté (EMONTAL)[19]. Ce corpus est thématiquement divers et comprend des périodiques de provenances diverses, comme des bulletins paroissiaux ou des journaux orientés politiquement (par exemple, communistes, centristes ou publiés par la Résistance). Les documents de notre corpus ont été collectés à partir des fonds régionaux Bourgogne et Franche-Comté de Gallica[20], le service numérique de la Bibliothèque Nationale de France (BnF). Parmi ces fonds, nous n'avons collecté que les documents pour lesquels la transcription OCR était disponible et fournie par la BnF. Celle-ci a été obtenue à l'aide du logiciel ABBYY FineReader. Ces documents sont enregistrés au format XML ALTO, qui contient le contenu textuel des documents ainsi qu'une description de leurs régions textuelles. Le tableau 1 montre la distribution des collections, documents, pages, lignes de textes et mots de notre corpus d'étude.

Le stockage des documents au format XML ALTO est une des principales difficultés pour le traitement de ce fond. En effet, si ce format contient les transcriptions OCR des documents ainsi qu'une description de leur mise en page, il n'est cependant pas adapté à l'application de méthodes de TAL. La diversité thématique du corpus est également une difficulté. En effet, bien que la langue des documents diffère peu de la langue moderne, les styles d'écriture ainsi que les registres de langue divergent suivant l'origine du document. Ainsi, les méthodes que nous appliquons pour traiter ce corpus doivent être suffisamment précises pour pouvoir s'adapter à cette hétérogénéité thématique et stylistique.

Notre objectif est de convertir les fichiers XML ALTO dans un format plus adapté à l'application de méthodes de TAL. Pour cela, nous avons mis en place une suite de pré-traitements pour annoter sémantiquement les lignes de texte et générer un nouveau document au format XML DocBook. Ce format à été choisi pour sa simplicité et sa modularité. Ces annotations pourraient cependant servir à générer des documents dans d'autres formats tels que XML TEI ou JSON. Nous appliquons les pré-traitements suivants : suppression de la césure, correction de l'OCR, analyse de la structure logique, division des paragraphes en phrases, division des documents en sections et entraînement de vecteurs de mots. Nous ne normalisons pas les variations orthographiques des documents en raison de leur proximité linguistique avec le français moderne. Nous entraînons ensuite des vecteurs de mots avant d'appliquer des annotations sémantiques aux document XML DocBook. Un exemple des résultats obtenus est montré par la figure 3. Une partie des résultats obtenus par l'application de ce cadre méthodologique a été publiée sous forme d'un jeu de données ouvert (Gutehrlé et Atanassova, 2021).

---

[19] http://tesniere.univ-fcomte.fr/projet-emontal/index.html
[20] https://gallica.bnf.fr/html/und/france/bourgogne-franche-comte

**Tableau 1** : *Descriptions des deux fonds composant le corpus EMONTAL*

| Corpus | Fond Franche-Comté | Fond Bourgogne | Total |
|---|---:|---:|---:|
| **Collections** | 46 | 118 | 164 |
| **Publications** | 2 650 | 6 120 | 8 770 |
| **Pages** | 255 908 | 709 250 | 965 158 |
| **Lignes** | 11 381 384 | 26 584 643 | 37 966 027 |
| **Mots** | 83 063 089 | 211 601 426 | 294 664 515 |

Nous procédons tout d'abord à une étape de suppression de la césure à l'aide d'un système de règles. Les césures sont représentées dans les fichiers XML ALTO par une balise HYP. Nous identifions donc ces balises afin de pouvoir reconstituer les mots séparés par une césure. La balise HYP est alors supprimée du document. Nous procédons ensuite à une étape de correction de l'OCR. L'entraînement et l'utilisation de modèles neuronaux pour la correction de l'OCR étant coûteuse informatiquement, nous avons fait le choix d'employer un système à base de règles. Ces dernières visent principalement à retirer les erreurs évidentes d'OCR tels que les textes illisibles et à identifier les mots nécessitants une correction orthographique. Une première règle consiste à supprimer tous mots composés d'une séquence de caractères non-alphanumériques. Les mots composés d'un seul caractère sont également supprimés, sauf les nombres et les exceptions en français tel que le « y ». Une seconde règle vient ensuite corriger les mots dont certains caractères se répètent plus de deux fois. Par exemple, « mercrediiii » est corrigé en « mercredi ». Enfin, une troisième règle emploie un dictionnaire français pour identifier les mots inconnus, hors noms propres, puis emploie le correcteur orthographique SymSpell[21] pour suggérer une correction. SymSpell reposant sur la distance d'édition (Levenshtein *et al.*, 1966) pour identifier des corrections possibles, nous rejetons tous candidats dont la distance d'édition est supérieure à 1.

Nous procédons ensuite à l'analyse de la structure logique des documents afin d'en identifier les titres, en-têtes, paragraphes. Les règles employées ainsi que leur évaluation sont présentés dans Atanassova et Gutehrlé (2022). Les paragraphes identifiés ainsi sont ensuite divisés en phrases au niveau des ponctuations (.?!). Des règles simples nous permettent de gérer des exceptions telles que « N.B. » ou « P.S. » et d'éviter une mauvaise division. De même, les titres identifiés nous permettent de diviser les documents en sections. Ainsi, une section est constituée par un titre ou groupe de titres, suivi de son texte jusqu'à rencontrer un autre titre. Enfin, après applications de ces étapes de pré-traitement, nous convertissons les documents XML ALTO au format XML DocBook.

---

[21] https://seekstorm.com/blog/1000x-spelling-correction/

Nous nous servons tout d'abord de ce corpus pour entraîner des vecteurs de mots à l'aide de la méthode *fastText* et de l'outil gensim. Nous appliquons ensuite les méthodes suivantes d'annotation sémantique : nous procédons tout d'abord à la reconnaissance d'entités nommées pour identifier les mentions de personnes et de lieux. Pour cela, nous avons entraîné un modèle dédié à partir du jeu de données HIPE-2020 (Ehrmann, Romanello, Flückiger, *et al.*, 2020), qui est consacré aux entités nommées dans les documents historiques. L'entraînement de ce modèle est réalisé à l'aide de l'outil spaCy et exploite les vecteurs de mots entraînés précédemment. Nous appliquons également la détection d'expressions temporelles pour identifier les expressions explicites comme implicites. Pour cela, nous employons l'outil HeidelTime. Enfin, nous employons l'outil Apache Solr[22] pour indexer nos documents et les rendre consultables par un moteur de recherche. Les documents sont indexés au niveau des termes présents mais également au niveau des annotations sémantiques, ce qui permet d'exprimer des requêtes très précises et de limiter le nombre de documents retournés.

**Figure 3** : *Fait divers extrait du journal communiste « Le Semeur » publié le 23 avril 1932 et extrait du XML Docbook correspondant, obtenu après application du cadre méthodologique*

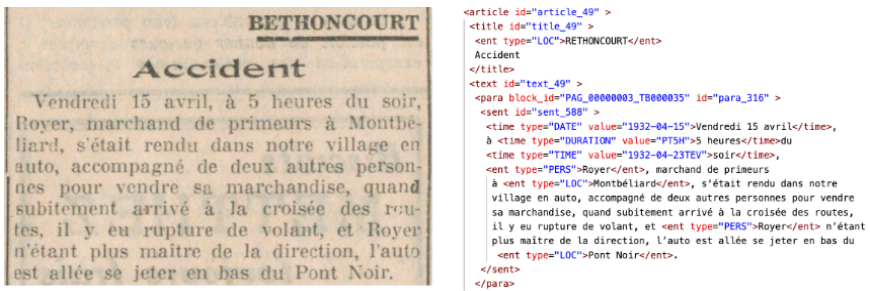

## 5. Conclusion

Dans le contexte des Humanités Numériques et face au nombre toujours plus important de documents d'archives disponibles aujourd'hui, il est nécessaire de concevoir des méthodologies pour structurer ces documents. Dans cet article, nous avons présenté un cadre méthodologique pour la construction de nouvelles interfaces de recherche qui reposent sur l'annotation sémantique d'un corpus d'étude de documents d'archives, dans le but d'en faciliter l'exploitation et la valorisation. Ce cadre se compose de trois grandes étapes qui sont le pré-traitement des documents, l'annotation de leur structure sémantique et l'augmentation des interfaces de recherche actuelles. Nous avons décrit les verrous méthodologiques et technologiques qui se présentent face à l'application de ce cadre, ainsi que leurs potentielles solutions. Enfin, nous avons présenté les outils de traitement existants ainsi qu'un cas pratique de l'application de ce cadre méthodologique.

---

[22] https://solr.apache.org/

Nous pensons que les traitements et outils proposés, ainsi que les verrous décrits dans cet article, peuvent guider tous projets visant à structurer de tels documents, dans le but de les exploiter et les valoriser. L'application complète de ce cadre à notre corpus d'étude fera l'objet de travaux futurs. De même, nous envisageons d'appliquer ce cadre à d'autres corpus de documents d'archives, mais également à d'autres domaines en sciences humaines, comme la littérature, afin d'identifier de potentiels nouveaux verrous et traitements nécessaires. Enfin, nous comptons également faire évaluer les interfaces obtenues par ce cadre méthodologique par des utilisateurs afin d'en estimer l'utilité dans la recherche d'information dans les documents d'archives.

## 6. Références

**Comprendre les archives : vers de nouvelles interfaces de recherche reposant sur l'annotation sémantique des documents**